\documentclass[a4paper,UKenglish,cleveref, autoref, thm-restate]{lipics-v2021}

\pdfoutput=1

\usepackage{extra}

\title{On Approximate Envy-Freeness for Indivisible Chores and Mixed Resources}

\titlerunning{Approximate EF for Indivisible Chores and Mixed Resources}

\author{Umang Bhaskar}{Tata Institute of Fundamental Research, India}{umang@tifr.res.in}{}{}
\author{A. R. Sricharan}{Chennai Mathematical Institute, India}{arsricharan@cmi.ac.in}{}{}
\author{Rohit Vaish}{Indian Institute of Technology Delhi, India}{rvaish@iitd.ac.in}{}{}

\authorrunning{U. Bhaskar, A. R. Sricharan, and R. Vaish}

\Copyright{Umang Bhaskar, A. R. Sricharan, and Rohit Vaish}



\relatedversiondetails[cite=BSV21approximateapprox]{Conference Version}{https://doi.org/10.4230/LIPIcs.APPROX/RANDOM.2021.1}

\acknowledgements{We thank the anonymous reviewers for helpful comments. UB acknowledges support from project no. RTI4001 of the Department of Atomic Energy, Government of India. RV acknowledges support from DST INSPIRE grant no. DST/INSPIRE/04/2020/000107. Part of this work was done while RV was affiliated with TIFR and was supported by Prof. R Narasimhan postdoctoral award and project no. RTI4001 of the Department of Atomic Energy, Government of India.}

\nolinenumbers

\EventEditors{Mary Wootters and Laura Sanit\`{a}}
\EventNoEds{2}
\EventLongTitle{Approximation, Randomization, and Combinatorial Optimization. Algorithms and Techniques (APPROX/RANDOM 2021)}
\EventShortTitle{\mbox{\scriptsize APPROX/RANDOM 2021}}
\EventAcronym{APPROX/RANDOM}
\EventYear{2021}
\EventDate{August 16--18, 2021}
\EventLocation{University of Washington, Seattle, Washington, US (Virtual Conference)}
\EventLogo{}
\SeriesVolume{207}
\ArticleNo{1}

\hideLIPIcs

\begin{document}

\maketitle

\begin{abstract}
We study the fair allocation of undesirable indivisible items, or \emph{chores}. While the case of desirable indivisible items (or \emph{goods}) is extensively studied, with many results known for different notions of fairness, less is known about the fair division of chores. We study envy-free allocation of chores and make three contributions. First, we show that determining the existence of an envy-free allocation is \NPC{} even in the simple case when agents have \emph{binary additive} valuations. Second, we provide a polynomial-time algorithm for computing an allocation that satisfies envy-freeness up to one chore (\EF{1}), correcting a claim in the existing literature. A modification of our algorithm can be used to compute an \EF{1} allocation for \emph{doubly monotone} instances (where each agent can partition the set of items into objective goods and objective chores). Our third result applies to a \emph{mixed resources} model consisting of indivisible items and a divisible, undesirable heterogeneous resource (i.e., a bad cake). We show that there always exists an allocation that satisfies envy-freeness for mixed resources (\EFM{}) in this setting, complementing a recent result of Bei et al.~\cite{BLL+21fair} for indivisible goods and divisible cake.
\end{abstract}

\section{Introduction}
\label{sec:Introduction}

The problem of fairly dividing a set of resources among agents is of central importance in various fields including economics, computer science, and political science. Such problems arise in many settings such as settling border disputes, assigning credit among contributing individuals, rent division, and distributing medical supplies such as vaccines~\cite{PSU+20fair}. The theoretical study of fair division has classically focused on \emph{divisible} resources (such as land or clean water), most prominently in the \emph{cake-cutting} literature~\cite{BT96fair,RW98cake,P15cake}; here, cake is a metaphor for a heterogeneous resource that can be fractionally allocated. A well-established concept of fairness in this setup is \emph{envy-freeness}~\cite{F67resource} which stipulates that no agent envies another, i.e., prefers the share of another agent to its own. An envy-free division of a divisible, desirable resource (i.e., a cake) is known to exist under general settings~\cite{S80cut,A87splitting,S99rental,AM16discrete}, and can be efficiently computed for a wide range of utility functions~\cite{CLP+11optimal,KLP13cut,AY14cake,BR20fair}.

By contrast, an envy-free solution can fail to exist when the goods are discrete or \emph{indivisible}; important examples include the assignment of course seats at universities~\cite{OSB10finding,BCK+17course} and the allocation of public housing units~\cite{BCH+20price}. This has motivated relaxations such as \emph{envy-freeness up to one good} (\EF{1}) where pairwise envy can be eliminated by removing some good from the envied bundle~\cite{LMM+04approximately,B11combinatorial}. The \EF{1} notion enjoys strong theoretical and practical motivation: On the theoretical side, there exist efficient algorithms for computing an \EF{1} allocation under general, monotone valuations~\cite{LMM+04approximately}. At the same time, \EF{1} has also found impressive practical appeal on the popular fair division website \emph{Spliddit}~\cite{GP15spliddit} and in course allocation applications~\cite{B11combinatorial,BCK+17course}.

Our focus in this work is on fair allocation of undesirable or negatively-valued indivisible items, also known as \emph{chores}. The chore division problem, introduced by Martin Gardner~\cite{G78aha}, models scenarios such as distribution of household tasks (e.g., cleaning, cooking, etc.) or the allocation of responsibilities for controlling carbon emissions among countries~\cite{T02fair}. For indivisible chores, too, an envy-free allocation could fail to exist, and one of our contributions is to show that determining the existence of such outcomes is \NPC{} even under highly restrictive settings (\Cref{thm:EF_NP-complete_Chores_Binary}). This negative result prompts us to explore the corresponding relaxation of \emph{envy-freeness up to one chore}, also denoted by \EF{1}, which addresses pairwise envy by removing some chore from the \emph{envious} agent's bundle.

At first glance, the chore division problem appears to be the `opposite' of the goods problem, and hence one might expect natural adaptation of algorithms that compute an \EF{1} allocation for goods to also work for chores. This, however, turns out to not be the case.

\emph{Goods vs chores}: Consider the well-known \emph{envy-cycle elimination} algorithm of Lipton et al.~\cite{LMM+04approximately} for computing an \EF{1} allocation of indivisible goods. Briefly, the algorithm works by iteratively assigning a good to an agent that is not envied by anyone else. The existence of such an agent is guaranteed by means of resolving cycles in the underlying \emph{envy graph}. (The envy graph of an allocation is a directed graph whose vertices correspond to the agents and there is an edge $(i,j)$ if agent $i$ envies agent $j$.) When adapted to the chores problem, the algorithm (\Cref{alg:Naive_EF1}) assigns a chore to a ``non-envious'' agent that has no outgoing edge in the envy graph. Contrary to an existing claim in the literature~\cite{ACI+18fair}, we observe that this algorithm could fail to find an \EF{1} allocation even when agents have additive valuations.\footnote{A recent work by B{\'e}rczi et al.~\cite{BBB+20envy} shows that this algorithm fails to find an \EF{1} allocation when agents have non-monotone and non-additive valuations. We show a stronger result, in that the failure in finding an \EF{1} allocation persists even when agents have additive, monotone nonincreasing valuations.}

\begin{algorithm}[t]
\DontPrintSemicolon
\KwIn{An instance $\langle N, M, \V \rangle$ with non-increasing valuations}
\KwOut{An allocation $A$}
\BlankLine
Initialize $A \leftarrow (\emptyset,\emptyset,\dots,\emptyset)$\;
\For{$c \in M$}{
    Choose a sink $i$ in the envy graph $G_A$\;
    Update $A_i \leftarrow A_i \cup \left\{ c \right\} $\;
    \While{ $G_A$ contains a directed cycle $C$ } {
        $A \leftarrow A^C$
    }
}
\KwRet $A$
\caption{Na\"ive envy-cycle elimination algorithm}
\label{alg:Naive_EF1}
\end{algorithm}

\newpage

\begin{figure}[t]
\centering
     \begin{subfigure}[b]{0.32\linewidth}
         \centering
         \begin{tikzpicture}
         	\tikzset{myredarrow/.style={
    decoration={markings,mark=at position 1 with {\arrow[scale=1.2,red]{>}}}, postaction={decorate}, shorten >=0.4pt}}
	     	\tikzset{myregulararrow/.style={
    decoration={markings,mark=at position 1 with {\arrow[scale=1.2]{>}}}, postaction={decorate}, shorten >=0.4pt}}
         	\draw (1,1.5) circle (7pt) node (a1) {$a_1$};
         	\node [above=of a1,yshift=-1cm] (bundle1) {\footnotesize{$\{c_1,c_4\}$}};
         	\draw (0,0) circle (7pt) node (a2) {$a_2$};
         	\node [below=of a2,yshift=1cm] (bundle2) {\footnotesize{$\{c_2,c_5\}$}};
         	\draw (2,0) circle (7pt) node (a3) {$a_3$};
         	\node [below=of a3,yshift=1cm] (bundle3) {\footnotesize{$\{c_3,c_6\}$}};
         	\draw[->,myredarrow,thick,red] (a2) to [bend right] (a3);
         	\draw[->,myregulararrow] (a3) to [bend right] (a2);
         	\draw[->,myredarrow,thick,red] (a1) to [out=0,in=60] (a3);
         	\draw[->,myredarrow,thick,red] (a3) to [bend right] (a1);
         \end{tikzpicture}
         \caption{Envy graph of $A$.}
         \label{fig:Envy_Graph_A}
     \end{subfigure}
     \hfill
     \begin{subfigure}[b]{0.32\linewidth}
         \centering
         \begin{tikzpicture}
         	\tikzset{myredarrow/.style={
    decoration={markings,mark=at position 1 with {\arrow[scale=1.2,red]{>}}}, postaction={decorate}, shorten >=0.4pt}}
	     	\tikzset{myregulararrow/.style={
    decoration={markings,mark=at position 1 with {\arrow[scale=1.2]{>}}}, postaction={decorate}, shorten >=0.4pt}}
         	\draw (1,1.5) circle (7pt) node (a1) {$a_1$};
			\node [above=of a1,yshift=-1cm] (bundle1) {\footnotesize{$\{c_3,c_6\}$}};
         	\draw (0,0) circle (7pt) node (a2) {$a_2$};
         	\node [below=of a2,yshift=1cm] (bundle2) {\footnotesize{$\{c_2,c_5\}$}};
         	\draw (2,0) circle (7pt) node (a3) {$a_3$};
         	\node [below=of a3,yshift=1cm] (bundle3) {\footnotesize{$\{c_1,c_4\}$}};
         	\draw[->,myredarrow,thick,red] (a2) to [bend left] (a1);
         \end{tikzpicture}
         \caption{Envy graph of $X$.}
         \label{fig:Envy_Graph_X}
     \end{subfigure}
     \hfill
     \begin{subfigure}[b]{0.32\linewidth}
         \centering
         \begin{tikzpicture}
         	\tikzset{myredarrow/.style={
    decoration={markings,mark=at position 1 with {\arrow[scale=1.2,red]{>}}}, postaction={decorate}, shorten >=0.4pt}}
	     	\tikzset{myregulararrow/.style={
    decoration={markings,mark=at position 1 with {\arrow[scale=1.2]{>}}}, postaction={decorate}, shorten >=0.4pt}}
         	\draw (1,1.5) circle (7pt) node (a1) {$a_1$};
         	\node [above=of a1,yshift=-1cm] (bundle1) {\footnotesize{$\{c_1,c_4\}$}};
         	\draw (0,0) circle (7pt) node (a2) {$a_2$};
         	\node [below=of a2,yshift=1cm] (bundle2) {\footnotesize{$\{c_3,c_6\}$}};
         	\draw (2,0) circle (7pt) node (a3) {$a_3$};
         	\node [below=of a3,yshift=1cm] (bundle3) {\footnotesize{$\{c_2,c_5\}$}};
         	\draw[->,myredarrow,thick,red] (a1) to [bend right] (a2);
         	\draw[->,myredarrow,thick,red] (a3) to [bend right] (a1);
         \end{tikzpicture}
         \caption{Envy graph of $Y$.}
         \label{fig:Envy_Graph_Y}
     \end{subfigure}
     \caption{Envy graphs of various allocations in \Cref{eg:Envy_Graph_Example_Intro}. A red edge points to the favorite (or most envied) bundle of an agent, while a black edge points to an envied (but not the most envied) bundle.}
\label{fig:Envy_Graph_Example}
\end{figure}
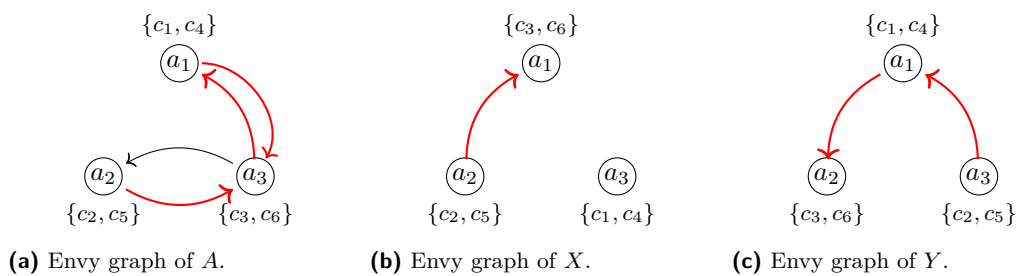

\begin{example}[\textbf{Envy-cycle elimination algorithm fails \EF{1} for additive chores}] Consider the following instance with six chores $c_1,\dots,c_6$ and three agents $a_1,a_2,a_3$ with additive valuations:

\begin{center}
 \begin{tabular}{c|c c c c c c}
& $c_1$ & $c_2$ & $c_3$ & $c_4$ & $c_5$ & $c_6$ \\ \hline
 $a_1$ & \circled{-1} & -4 & -2 & \circled{-3} & 0 & -1 \\
$a_2$ & -2 & \circled{-1} & -2 & -2 & \circled{-3} & -1\\
$a_3$ & -1 & -3 & \circled{-1} & -1 & -3 & \circled{-10}\\ \end{tabular}
\end{center}

Suppose the algorithm considers the chores in the increasing order of their indices (i.e., $c_1$, $c_2$, $\ldots$), and breaks ties among agents in favor of $a_1$ and then $a_2$. No directed cycles appear at any intermediate step during the execution of the algorithm on the above instance. The resulting allocation, say $A$, is given by $A_1 = \{c_1,c_4\}$, $A_2 = \{c_2,c_5\}$, and $A_3 = \{c_3,c_6\}$ (shown as circled entries in the above table). Notice that $A$ is \EF{1} and its envy graph is as shown in \Cref{fig:Envy_Graph_A}.

Each node in the envy graph of $A$ has an outgoing edge (\Cref{fig:Envy_Graph_A}). Therefore, if the algorithm were to allocate another chore after this, it would have to resolve either the cycle $\{a_1,a_3\}$ or the cycle $\{a_2,a_3\}$. Let $X$ and $Y$ denote the allocations obtained by resolving the cycles $\{a_1,a_3\}$ and $\{a_2,a_3\}$, respectively (the corresponding envy graphs are shown in \Cref{fig:Envy_Graph_X,fig:Envy_Graph_Y}). Although both envy graphs are \emph{acyclic} (and thus admit a ``sink'' agent), only the allocation $X$ satisfies \EF{1}; in particular, the pair $\{a_1,a_3\}$ violates \EF{1} for $Y$.
\label{eg:Envy_Graph_Example_Intro}
\end{example}

The above example highlights an important contrast between indivisible goods and chores: For goods, resolving arbitrary envy cycles preserves \EF{1}, whereas for chores, the choice of which envy cycle is resolved matters.
This is because when evaluating \EF{1} for chores, a chore is removed from the envious agent's bundle. In the envy-cycle resolution step, if a cycle is chosen without caution, then it is possible for an agent to acquire a bundle that, although strictly more preferable, contains no chore that is large enough to compensate for the envy on its own.

A key insight of our work is that there always exists a specific envy cycle---the \emph{top-trading envy cycle}---that can be resolved to compute an \EF{1} allocation of chores. Our algorithm computes \EF{1} allocations for \emph{monotone} valuations, and thus provides an analogue of the result of Lipton et al.~\cite{LMM+04approximately} for the chores setting. Furthermore, a simple modification of our algorithm computes an \EF{1} allocation for \emph{doubly monotone} instances (\Cref{thm:Doubly_Monotone_EF1}), where each agent can partition the items into `objective goods' and `objective chores', i.e., items with non-negative and negative marginal utility, respectively, for the agent~\cite{ACI+18fair}. This class has also been referred to as \emph{itemwise monotone} in the literature~\cite{CL20fairness}.

Motivated by this positive observation, we study a \emph{mixed} model, consisting of both divisible as well as indivisible resources. This is a natural model in many settings, e.g., dividing an inheritance that consists of both property and money, or the simultaneous division of chores and rent among housemates. Although the use of payments in fair allocation of indivisible resources has been explored in several works~\cite{M87fair,ADG91fair,A95derivation,S99rental,K00algorithm,MPR02envy,HRS02bidding,HS19fair,A21achieving,BDN+20one,CI20computing}, the most general formulation of a model with mixed resources, in our knowledge, is due to Bei et al.~\cite{BLL+21fair} who study combined allocation of a divisible \emph{heterogeneous} resource (i.e., a cake) and a set of indivisible goods. This model and its variants are the focus of our work.

Generalizing the set of resources calls for revising the fairness benchmark. While exact envy-freeness still remains out of reach in the mixed model, \EF{1} can be ``too permissive'' when only the divisible resource is present. Bei et al.~\cite{BLL+21fair} propose a fairness concept called \emph{envy-freeness for mixed goods} (\EFM{}) for indivisible goods and divisible cake, which evaluates fairness with respect to \EF1{} if the envied bundle only contains indivisible goods, but switches to exact envy-freeness if the envied agent is allocated any cake.
They show that an \EFM{} allocation always exists for a mixed instance when agents have additive valuations within as well as across resource types. We note that neither the algorithm of Bei et al.~\cite{BLL+21fair} nor its analysis crucially depends on the valuations for the indivisible goods being additive; in fact, their results extend to monotone valuations for the indivisible goods.

We consider the problem of envy-free allocation in mixed instances with doubly monotone indivisible items and bad cake. We extend the definition of \EFM{} naturally to this model, and show that in this model as well, an \EFM{} allocation always exists~(\Cref{thm:EFM_DM_Bad_Cake_Termination}). Our work thus extends the results of Bei et al. in two ways --- allowing for bad cake as well as doubly monotone indivisible items.

We also study a mixed model with indivisible chores and good cake. This turns out to be quite challenging, because unlike previous cases, one cannot start with an arbitrary \EF1{} allocation of the indivisible items and then allocate cake to obtain an \EFM{} allocation. We however show the existence of an \EFM{} allocation for two special cases in this model: when each agent has the same ranking over the chores~(\Cref{thm:EFM_Chores_Cake_Identical_Rankings}), and when the number of chores is at most one more than the number of agents~(\Cref{thm:EFM_Chores_Cake_Small_No_of_Items}).

\begin{table*}[t]
\centering
\scriptsize
\renewcommand{\arraystretch}{1.7}
\begin{tabular}{|c|c|c|}
\hline
\diagbox{\textbf{Indivisible}}{\textbf{Divisible}} & \textbf{Cake} & \textbf{Bad Cake} \\\hline
&&\\[-1.5em]
\textbf{Goods} & \checkmark~\cite{BLL+21fair} & \multirow{3}{*}{\checkmark~(\Cref{thm:EFM_DM_Bad_Cake_Termination}) }\\[0.2em]
\cline{1-2}
&&\\[-1.5em]
\multirow{2}{*}{\textbf{Chores}} & \checkmark for identical rankings~(\Cref{thm:EFM_Chores_Cake_Identical_Rankings}) & \\
& \checkmark for $n+1$ items~(\Cref{thm:EFM_Chores_Cake_Small_No_of_Items}) & \\[0.2em]
\hline
\end{tabular}
\vspace{3mm}
\caption{\EFM{} existence results for different combinations of indivisible and divisible resources. A \checkmark indicates that \EFM{} exists in that setting.}
\label{tab:EFM_Results}
\end{table*}

\paragraph*{Our Contributions}

\begin{enumerate}
    \item We first show that determining whether an envy-free allocation of chores exists is strongly \NPC{}, even in the highly restricted setting when agents have \emph{binary additive} valuations, i.e., when for all agents $i \in [n]$ and items $j \in [m]$, $v_{i,j} \in \{-1,0\}$~(\Cref{thm:EF_NP-complete_Chores_Binary}). The analogous problem for indivisible goods with binary valuations is already known to be \NPC{}~\cite{AGM+15fair,HSV+20fair}.

    \item When the fairness goal is relaxed to envy-freeness up to one chore (\EF{1}), we establish efficient computation for instances with chores (\Cref{thm:Monotone_Chores}), and instances with \emph{doubly monotone} valuations, when each agent $i$ can partition the set of items into goods $G_i$ (which always have nonnegative marginal value) and chores $C_i$ (which always have nonpositive marginal value) (\Cref{thm:Doubly_Monotone_EF1}).

    \item For a mixed instance consisting of doubly monotone indivisible items and bad cake, we show the existence of an allocation that satisfies the stronger fairness guarantee called \emph{envy-freeness up to a mixed item} (\EFM{}) (\Cref{thm:EFM_DM_Bad_Cake_Termination}). Our result uses our previous theorem for indivisible chores as well as the framework of Bei et al.~\cite{BLL+21fair} for the allocation of the divisible item. This complements the result of Bei et al.~\cite{BLL+21fair} by showing existence of \EFM{} allocations for mixed instances consisting of both desirable and undesirable items.

    \item Lastly, for a mixed instance consisting of indivisible chores and (good) cake (see \Cref{subsec:EFM_Special_Cases_Chores_and_Cake}), we show the existence of an \EFM{} allocation in two special cases: when each agent has the same preference ranking over the set of items (\Cref{thm:EFM_Chores_Cake_Identical_Rankings}), and when the number of items is at most one more than the number of agents~(\Cref{thm:EFM_Chores_Cake_Small_No_of_Items}).
    \end{enumerate}

    Our results for mixed instances are summarized in \Cref{tab:EFM_Results}.

\section{Related Work}

As mentioned, fair division has been classically studied for \emph{divisible} resources. For a \emph{heterogeneous}, desirable resource (i.e., a cake), the existence of envy-free solutions is known under mild assumptions~\cite{S80cut,A87splitting,S99rental,AM16discrete}. In addition, efficient algorithms are known for computing $\eps$-envy-free divisions~\cite{P15cake} and envy-free divisions under restricted preferences~\cite{CLP+11optimal,KLP13cut,AY14cake,BR20fair}. For an undesirable heterogeneous resource (a bad cake), too, the existence of an envy-free division is known~\cite{PS09nperson}, along with a discrete and bounded procedure for finding such a division~\cite{DFH+18envy}. For the case of non-monotone cake (i.e., a real-valued divisible heterogeneous resource), the existence of envy-free outcomes has been shown for specific numbers of agents~\cite{S18fairly,MZ19envy,AK19envy,AK20equipartition}.

Turning to the \emph{indivisible} setting, we note that the sweeping result of Lipton et al.~\cite{LMM+04approximately} on \EF{1} for indivisible \emph{goods} has inspired considerable work on establishing stronger existence and computation guarantees in conjunction with other well-studied economic properties~\cite{CKM+19unreasonable,BKV18Finding,BKV18greedy,FSV+19equitable,BCI+20finding,CGM21fair,AMN20multiple,FSV20best}. The case of \emph{indivisible chores} has been similarly well studied for a variety of solution concepts such as maximin fair share~\cite{ACL19weighted,ARS+17algorithms,ALW19strategyproof,HL21algorithmic}, equitability~\cite{FSV+20equitable,A20jealousy}, competitive equilibria with general incomes~\cite{S20competitive}, and envy-freeness~\cite{A18almost,BS19algorithms,FSV20best}.

Aziz et al.~\cite{ACI+18fair,ACI+19fair} study a model containing both indivisible goods and chores, wherein \emph{envy-freeness up to an item} (\EF{1}) entails that pairwise envy is bounded by the removal of some good from the envied bundle or some chore from the envious agent's bundle. They show that a variant of the classical round-robin algorithm computes an \EF{1} allocation under additive utilities, and also claim that a variant of the envy-cycle elimination algorithm~\cite{LMM+04approximately} returns such allocations for doubly monotone instances (we revisit the latter claim in \Cref{eg:Envy_Graph_Example_Intro}).  Other fairness notions such as approximate proportionality~\cite{ACI+18fair,ACI+19fair,AMS20polynomial}, maximin fair share~\cite{KMT21approximating}, approximate jealousy-freeness~\cite{A20jealousy}, and weaker versions of \EF{1}~\cite{FSV20best} have also been studied in this model.

Finally, we note that the model with \emph{mixed resources} comprising of both indivisible and (heterogeneous) divisible parts has been recently formalized by Bei et al.~\cite{BLL+21fair}, although a special case of their model where the divisible resource is homogenous and desirable (e.g., money) has been extensively studied~\cite{M87fair,ADG91fair,A95derivation,S99rental,K00algorithm,MPR02envy,HRS02bidding,HS19fair,A21achieving,BDN+20one,CI20computing}. Bei et al.~\cite{BLL+21fair} showed that when there are indivisible goods and a divisible cake, an allocation satisfying \emph{envy-freness for mixed goods} (\EFM{}) always exists. Subsequent work considers the maximin fairness notion in the mixed model~\cite{BLL+21maximin}.

\section{Preliminaries}
\label{sec:Preliminaries}

We consider two kinds of instances: one with purely indivisible items and the other with a mixture of divisible and indivisible items. We will present the preliminaries for instances with purely indivisible items in this section, and defer details for the mixed resources model to \Cref{sec:Approximate_EF_Mixed_Resources}.

\paragraph*{Problem instance:}
An \emph{instance} $\langle N, M, \V \rangle$ of the fair division problem is defined by a set $N$ of $n \in \N$ \emph{agents}, a set $M$ of $m \in \N$ \emph{indivisible items}, and a \emph{valuation profile} $\V = \{v_1,v_2,\dots,v_n\}$ that specifies the preferences of every agent $i \in N$ over each subset of the items in $M$ via a \emph{valuation function} $v_i: 2^{M} \rightarrow \mathbb{R}$.

\paragraph*{Marginal valuations:} For any agent $i \in N$ and any set of items $S \subseteq M$, the \emph{marginal valuation} of the set $T \subseteq M \setminus S$ is given by $v_i(T | S) \coloneqq v_i(S \cup T) - v_i(S)$. When the set $T$ is a singleton (say $T = \{j\}$), we will write $v_i(j|S)$ instead of $v_i(\{j\}|S)$ for simplicity.

\paragraph*{Goods and chores:} Given an agent $i \in N$ and an item $j \in M$, we say that $j$ is a \emph{good} for agent $i$ if for every subset $S \subseteq M \setminus \{j\}$, $v_i(j|S) \geq 0$. We say that $j$ is a \emph{chore} for agent $i$ if for every subset $S \subseteq M \setminus \{j\}$, $v_i(j|S) \le 0$, with one of the inequalities strict. Note that for general valuations, an item may neither be a good nor a chore for an agent.

\paragraph*{Doubly monotone instances:} An instance is said to be \emph{doubly monotone} if for each agent, each item is either a good or a chore. That is, each agent $i$ can partition the items as $M = G_i \uplus C_i$, where $G_i$ are her goods, and $C_i$ are her chores. Note that an item may be a good for one agent and a chore for another.

\paragraph*{Monotone instances:} A valuation function $v$ is \emph{monotone non-decreasing} if for any sets $S \subseteq T \subseteq M$, we have $v(T) \geq v(S)$, and \emph{monotone non-increasing} if for any sets $S \subseteq T \subseteq M$, we have $v(S) \geq v(T)$. A \emph{monotone goods} instance is one where all the agents have monotone non-decreasing valuations, and a \emph{monotone chores} instance is one where all the agents have monotone non-increasing valuations. We refer to such an instance as a \emph{monotone} if it is clear from context whether we are working with goods or chores.

\paragraph*{Additive valuations:} A well-studied subclass of monotone valuations is that of \emph{additive valuations}, wherein an agent's
value of any subset of items is equal to the sum of the values of individual items in the set, i.e., for any agent $i \in N$ and any set of items $S \subseteq M$, $v_i(S) \coloneqq \sum_{j \in S} v_i(\{j\})$, where we assume that $v_i(\emptyset) = 0$. For simplicity, we will write $v_i(j)$ or $v_{i,j}$ to denote $v_i(\left\{ j \right\})$.

\paragraph*{Allocation:}
An \emph{allocation} $A \coloneqq (A_1,\dots,A_n)$ is an $n$-partition of a subset of the set of items $M$, where $A_i \subseteq M$ is the \emph{bundle} allocated to the agent $i$ (note that $A_i$ can be empty). An allocation is said to be \emph{complete} if it assigns all items in $M$, and is called \emph{partial} otherwise.

\paragraph*{Envy graph:}
The \emph{envy graph} $G_A$ of an allocation $A$ is a directed graph on the vertex set $N$ with a directed edge from agent $i$ to agent $k$ if $v_i(A_k) > v_i(A_i)$, i.e., if agent $i$ prefers the bundle $A_k$ over the bundle $A_i$.

\paragraph*{Top-trading envy graph:}
The \emph{top-trading envy graph} $T_A$ of an allocation $A$ is a subgraph of its envy graph $G_A$ with a directed edge from agent $i$ to agent $k$ if $v_i(A_k) = \max_{j \in N} v_i(A_j)$ and $v_i(A_k) > v_i(A_i)$, i.e., if agent $i$ envies agent $k$ and $A_k$ is the most preferred bundle for agent $i$.

\paragraph*{Cycle-swapped allocation:}
Given an allocation $A$ and a directed cycle $C$ in an envy graph or a top-trading envy graph, the \emph{cycle-swapped allocation} $A^C$ is obtained by reallocating bundles backwards along the cycle. For each agent $i$ in the cycle, define $i^+$ as the agent that she is pointing to in $C$. Then, $A_i^C = A_{i^+}$ if $i \in C$, otherwise $A_i^C = A_i$.

\paragraph*{Envy-freeness and its relaxations:}
An allocation $A$ is said to be
\begin{itemize}
	\item \emph{envy-free} (\EF{}) if for every pair of agents $i,k \in N$, we have $v_i(A_i) \geq v_i(A_k)$, and

	\item \emph{envy-free up to one item} (\EF{1}) if for every pair of agents $i,k \in N$ such that $A_i \cup A_k \neq \emptyset$, there exists an item $j \in A_i \cup A_k$ such that $v_i(A_i \setminus \{j\}) \geq v_i(A_k \setminus \{j\})$.
\end{itemize}

\section{Envy-Freeness for Binary Valued Chores}
\label{sec:EF_Binary_Valued}

Our first result shows that determining the existence of an envy-free allocation is \NPC{} even when agents have binary valuations, i.e., when, for all agents $i \in N$ and items $j \in M$, $v_{i,j} \in \{-1,0\}$ (\Cref{thm:EF_NP-complete_Chores_Binary}). If agent valuations are not binary-valued, but are identical, the problem is still (weakly) \NPC{} via a straightforward reduction from \Partition{}. By contrast, our result establishes strong \NPC{}ness.

\begin{restatable}{theorem}{NPhardness}
Determining whether a given chores instance admits an envy-free allocation is \NPC{} even for binary utilities.
\label{thm:EF_NP-complete_Chores_Binary}
\end{restatable}

The proof of \Cref{thm:EF_NP-complete_Chores_Binary} can be found in \Cref{subsec:Proof_NPC_Chores_Binary}.

\section{\EF{1} For Doubly Monotone Instances}
\label{sec:SEF1}

In light of the intractability result in the previous section, we will now explore whether one can achieve approximate envy-freeness (specifically, \EF{1}) for indivisible chores. To this end, we note that the well-known round-robin algorithm (where, in each round, agents take turns in picking their favorite available chore) computes an \EF{1} allocation when agents have \emph{additive} valuations~\cite{ACI+19fair}. In the following, we will provide an algorithm for computing an \EF{1} allocation for the much more general class of \emph{monotone valuations}. Thus, our result establishes the analogue of the result of Lipton et al.~\cite{LMM+04approximately} from the goods-only model for indivisible chores.

\subsection{An Algorithm for Monotone Chores}
\label{subsec:Monotone_Chores}

As previously mentioned, the algorithm of Lipton et al.~\cite{LMM+04approximately} computes an \EF{1} allocation for indivisible goods under monotone valuations. Recall that the algorithm works by assigning, at each step, an unassigned good to an agent who is not envied by anyone else (such an agent is a ``source'' agent in the underlying envy graph). The existence of such an agent is guaranteed by resolving arbitrary envy cycles in the envy graph until it becomes acyclic. Importantly, resolving an arbitrary envy cycle preserves \EF{1}.

To design an \EF{1} algorithm for indivisible chores, prior work~\cite{ACI+18fair,ACI+19fair} has proposed the following natural adaptation of this algorithm (see \Cref{alg:Naive_EF1}): Instead of a ``source'' agent, an unassigned chore is now allocated to a ``sink'' (i.e., non-envious) agent in the envy graph. The existence of such an agent is once again guaranteed by means of resolving envy cycles. However, as noted in \Cref{eg:Envy_Graph_Example_Intro}, resolving \emph{arbitrary} envy cycles could destroy the \EF{1} property.

\enlargethispage{0.4cm}

To address this gap, we propose to resolve a specific envy cycle that we call the \emph{top-trading envy cycle}. (The nomenclature is inspired from the celebrated top-trading cycle algorithm~\cite{SS74cores} for finding a core-stable allocation that involves cyclic swaps of the most preferred objects.) Specifically, given a partial allocation $A$, we consider a subgraph of the envy graph $G_A$ that we call the \emph{top-trading envy graph} $T_A$ whose vertices denote the agents, and an edge $(i,k)$ denotes that agent $i$'s (weakly) most preferred bundle is $A_k$.

It is easy to observe that if the envy graph does not have a sink, then the top-trading envy graph $T_A$ has a cycle (\Cref{lem:TTCExistPolyTime}). Thus, resolving top-trading envy cycles (instead of arbitrary envy cycles) also guarantees the existence of a sink agent in the envy graph. More importantly, though, resolving a top-trading envy cycle preserves \EF{1}. Indeed, every agent involved in the top-trading exchange receives its most preferred bundle after the swap, and therefore does not envy anyone else in the next round. The resulting algorithm is presented in Algorithm~\ref{alg:TTC_EF1}.

\begin{algorithm}[t]
\DontPrintSemicolon
\KwIn{An instance $\langle N, M, \V \rangle$ with non-increasing valuations}
\KwOut{An allocation $A$}
\BlankLine
Initialize $A \leftarrow (\emptyset,\emptyset,\dots,\emptyset)$\;
\For{$c \in M$} {
    \If{there is no sink in $G_A$} {
        $C \leftarrow $ any cycle in  $T_A$ \Comment*[r]{\footnotesize{if $G_A$ has no sink, then $T_A$ must have a cycle (\Cref{lem:TTCExistPolyTime})}}
          $A \leftarrow A^C$
      }
    Choose a sink $k$ in the graph $G_A$\;
    Update $A_k \leftarrow A_k \cup \left\{ c \right\} $
}
\KwRet $A$
\caption{Top-trading envy-cycle elimination algorithm}
\label{alg:TTC_EF1}
\end{algorithm}

\begin{restatable}{theorem}{Monotone_Chores}
\label{thm:Monotone_Chores}
For a monotone instance with indivisible chores, Algorithm~\ref{alg:TTC_EF1} returns an \EF{1} allocation.
\end{restatable}

In \Cref{subsec:Doubly_monotone}, we will discuss a more general result (\Cref{thm:Doubly_Monotone_EF1}) that extends the top-trading envy-cycle elimination algorithm to \emph{doubly monotone} instances containing both indivisible goods as well as indivisible chores.

\subsection{An Algorithm for Doubly Monotone Instances}
\label{subsec:Doubly_monotone}

For a doubly monotone instance with indivisible items, we now give an algorithm (Algorithm~\ref{alg:SEF1}) that returns an \EF{1} allocation. The algorithm runs in two phases. The first phase is for all the items that are a good for at least one agent. For these items, we run the envy-cycle elimination algorithm of Lipton et al.~\cite{LMM+04approximately}, but restricted to the subgraph of agents who consider the item a good. In the second phase, we allocate items that are chores to everybody by running the top-trading envy-cycle elimination algorithm. For a monotone chores-only instance, we recover Algorithm~\ref{alg:TTC_EF1} as a special case of Algorithm~\ref{alg:SEF1}.

\begin{algorithm}[t]
\DontPrintSemicolon
\KwIn{An instance $\langle N, M, \V, \{G_i\}, \{C_i\} \rangle$ with indivisible items and doubly monotone valuations, where $G_i$ and $C_i$ are the set of goods and chores for agent $i$, respectively}
\KwOut{An allocation $A$}
$A \leftarrow (\emptyset, \emptyset, \ldots, \emptyset)$\;
\tcp{Goods Phase}
\For{each item $g \in \cup_i G_i$} {
    $V^g = \left\{ i \in N \mid g \in G_i \right\}$ \Comment*[r]{\footnotesize{$V^g$ contains all agents for whom g is a good}}
    $G_A^g =$ the envy graph $G_A$ restricted to the vertices $V^g$\;
    Choose a source $k$ in the graph $G_A^g$\;
    Update $A_{k} \leftarrow A_{k} \cup \left\{ g \right\} $\;
    \While{$G_A$ contains a directed cycle $C$} {
        $A \leftarrow A^C$
    }
}
\tcp{Chores Phase}
\For{each item $c \in \cap_i C_i$} {
    \If{there is no sink in $G_A$\label{algline:Chores_Phase_1}} {
        $C \leftarrow $ any cycle in  $T_A$ \Comment*[r]{\footnotesize{if $G_A$ has no sink, then $T_A$ must have a cycle}}
          $A \leftarrow A^C$
      }
    Choose a sink $k$ in the graph $G_A$\;
    Update $A_k \leftarrow A_k \cup \left\{ c \right\} $
}
\KwRet $A$
\caption{An \EF{1} algorithm for doubly monotone indivisible instances}
\label{alg:SEF1}
\end{algorithm}

\begin{restatable}{theorem}{SEF}
\label{thm:Doubly_Monotone_EF1}
For a doubly monotone instance with indivisible items, Algorithm~\ref{alg:SEF1} returns an \EF{1} allocation.
\end{restatable}

We first provide a brief sketch of the proof: At each step, we maintain the invariant that the partial allocation maintained by the algorithm is \EF{1}. This is certainly true for the goods phase, where any envy created from agent $i$ to agent $j$ can always be eliminated by removing a good $g \in A_j$ (however, unlike in the envy-cycle cancellation for goods-only instances~\cite{LMM+04approximately}, the eliminated item may not be the most recently added one since such an item could be a chore for an envious agent). In the chores phase, any new envy created by adding a chore can be removed by dropping the newly added chore. If we resolve top-trading envy cycles, then none of the agents within the cycle envy any of the agents outside it, since they now have their most preferred bundle. For any agent $i$ outside the cycle, any envy can be removed by either removing a chore from $i$ or a good from the envied bundle, since $i$'s allocation is unchanged and the bundles remain unbroken.

\begin{restatable}{lemma}{SEFGoods}
\label{lem:SEF1_Goods}
After every step of the goods phase, the partial allocation remains \EF{1}. Further, the goods phase terminates in polynomial time.
\end{restatable}

The proof of \Cref{lem:SEF1_Goods} closely follows the arguments of Lipton et al.~\cite{LMM+04approximately}; for completeness, we present a self-contained proof in \Cref{subsec:Proof_SEF1_Goods}. We will now consider the chores phase of the algorithm, and show that if there is no sink in the envy graph $G_A$, then there is a cycle in the top-trading envy graph $T_A$.

\begin{restatable}{lemma}{TTCExistPolyTime}
\label{lem:TTCExistPolyTime}
Let $A$ be a partial allocation whose envy graph $G_A$  does not have a sink. Then, the top-trading envy graph $T_A$ must have a cycle. Furthermore, such a cycle can be found in polynomial time.
\end{restatable}
\begin{proof}
Since $G_A$ has no sinks, every vertex in $G_A$ has outdegree at least one. Thus for all agents $i$, $i \not\in \arg \max_k v_i(A_k)$. So even in the top-trading envy graph $T_A$, each vertex has outdegree at least one. We start at an arbitrary agent and follow an outgoing edge from each successive agent. This gives us a cycle in $T_A$. It is easy to see that finding the cycle takes only polynomial time since we encounter each vertex at most once.
\end{proof}

We now show that resolving a cycle in the top-trading envy graph $T_A$ gives an allocation that necessarily has a sink (the existence of such a cycle in $T_A$ is given by \Cref{lem:TTCExistPolyTime}).

\begin{restatable}{lemma}{ResolveTTCSinks}
\label{lem:ResolveTTCSinks}
Let $A$ be a partial allocation whose top-trading envy graph $T_A$ contains a cycle. Let $A'$ denote the allocation obtained by resolving an arbitrary cycle in $T_A$. Then the envy graph $G_{A'}$ of the allocation $A'$ must have a sink.
\end{restatable}
\begin{proof}
Note that each agent points to its favorite bundle in $T_A$. Thus after resolving a cycle in $T_A$, all agents who participated in the cycle-swap now have their most preferred bundle in $A'$ and do not envy any other agent. These agents are sinks in the graph $G_{A'}$.
\end{proof}

To show that the partial allocation remains \EF{1} throughout the chores phase, we use Lemmas \ref{lem:SEF1_Chores_Add} and \ref{lem:SEF1_Chores_TTC_Resolve}.

\begin{restatable}{lemma}{SEFoneChoresAdd}
\label{lem:SEF1_Chores_Add}
In the chores phase, adding a new chore to the allocation (Line 13-14) preserves \EF{1}.
\end{restatable}
\begin{proof}
Suppose at time step $t$, the algorithm assigns a new chore (Line 13-14). Suppose before time step $t$, our allocation $A$ was \EF{1}, and the allocation after time step $t$ is $A'$. We show that $A'$ is \EF{1} as well. A sink exists in the envy graph $G_A$ at time step $t$, either because there were no top-trading envy cycles when we entered the loop (at Line 9) which implies the existence of a sink, or because we resolved a top-trading envy cycle $C$ in the previous time step $t-1$ (Lines 10-12), in which case all the agents who were a part of the resolved top-trading envy cycle do not envy anyone after the cycle swap, and are sinks in the envy graph $G_A$.

Then after time $t$, the allocation $A'$ will be $A_k' = A_k \cup \{c\}$, and  $A_j' = A_j$ for all $j\neq k$, where $k$ is a sink in $G_A$. Pick two agents $i$ and $j$ such that $i$ envies $j$ in $A'$. If $i$ did not envy $j$ in $A$, then clearly $i=k$. In this case, removing $c$ from $A_i$ removes $i $'s envy. Suppose $i$ envied $j$ in $A$ as well, and the envy was eliminated by removing $o \in A_i \cup A_j$. Then $i \neq k$ since $k$ was a sink in the graph $G_A$, and so $v_i(A_i) = v_i(A_i')$. If $o \in A_i$, then $v_i(A_i' \setminus \{o\}) \ge v_i(A_j) \ge v_i(A_j')$. If $o \in A_j$, then $v_i(A_i') \ge v_i(A_j \setminus \{o\}) \ge v_i(A_j \cup \{c\} \setminus \{o\})$, since $c$ is a chore for all agents.
\end{proof}

\begin{restatable}{lemma}{SEFoneChoresTTCResolve}
\label{lem:SEF1_Chores_TTC_Resolve}
In the chores phase, resolving a top-trading envy cycle (Lines 10-12) preserves \EF{1}.
\end{restatable}
\begin{proof}
Suppose at time step $t$, the algorithm resolves a top-trading envy cycle (Line 10-12). Let $A$ be the allocation before time $t$, $C$ be the cycle along which the swap happens, and $A' = A^C$ the allocation obtained by swapping backwards along the cycle. Pick two agents $i$ and $j$ such that $i$ envies $j$ in $A'$. Since every agent in the cycle obtains their favorite bundle, $i \not\in C$. Thus $A_i = A_i'$. Let $j'$ be the agent such that $A_j' = A_{j'}$. Since $v_i(A_i') = v_i(A_i)$, $i$ envied $j'$ before the swap which could be eliminated by removing $o \in A_i \cup A_{j'}$. If $o \in A_i$, then $v_i(A_i' \setminus \{o\}) \ge v_i(A_j')$. If $o \in A_{j'}$, then $v_i(A_i') \ge v_i(A_j' \setminus \{o\})$. Thus removing $o \in A_i \cup A_{j'}$ eliminates the envy in $A'$.
\end{proof}

By \Cref{lem:SEF1_Goods}, the allocation at the beginning of the chores phase is \EF{1}. At every time step $t$ of the chores phase, the algorithm either assigns a chore to a sink agent or resolves a top-trading envy cycle. Thus \Cref{lem:SEF1_Chores_Add} and \Cref{lem:SEF1_Chores_TTC_Resolve} together show that the allocation remains \EF{1} throughout the chores phase. By \Cref{lem:TTCExistPolyTime}, finding a cycle in $T_A$ takes only polynomial time. Since the while-loop executes only once for each chore, the chores phase terminates in polynomial time. This gives us the following lemma:

\begin{restatable}{lemma}{SEF1Chores}
\label{lem:SEF1_Chores}
At every step of the chores phase, the allocation remains \EF{1}, and the chores phase terminates in polynomial time.
\end{restatable}

The proof of \Cref{thm:Doubly_Monotone_EF1} follows immediately, since by \Cref{lem:SEF1_Chores} the allocation at the end of the chores phase is \EF{1}. Thus Algorithm~\ref{alg:SEF1} returns an \EF{1} allocation for a doubly monotone instance. Specialized to instances with only chores, we obtain \Cref{thm:Monotone_Chores} as a corollary.

\section{Approximate Envy Freeness for Mixed Resources}
\label{sec:Approximate_EF_Mixed_Resources}

We will now describe the setting with \emph{mixed resources} consisting of both divisible and indivisible parts. This model was recently studied by Bei et al.~\cite{BLL+21fair}, who introduced the notion of \emph{envy-freeness for mixed goods} (\EFM{}) in the context of a model consisting of indivisible goods and a divisible cake. We generalize this notion to a setting with both goods and chores.

\subsection{Preliminaries for Instances with Divisible and Indivisible Resources}
\label{subsec:Preliminaries_Mixed}

\paragraph*{Mixed instance:}
A mixed instance $\langle N, M, \V, \C, \F \rangle$ is defined by a set of $n$ agents, $m$ indivisible items, a valuation profile $\V$ (over the indivisible items), a divisible resource $\C$ represented by the interval $[0,1]$, and a family $\F$ of \emph{density functions} over the divisible resource. The valuations for the indivisible items are as described in \Cref{sec:Preliminaries}. For the divisible resource, each agent has a \emph{density function} $f_i: [0,1] \rightarrow \mathbb{R}$ such that for any measurable subset $S \subset [0,1]$, agent $i$ values it at $v_i(S) \coloneqq \int_S f_i(x) dx$. When the density function is non-negative for every agent (i.e., for all $i \in N$, $f_i: [0,1] \rightarrow \mathbb{R}_{\geq 0}$), we will call the divisible resource a ``cake'', and for non-positive densities (i.e., for all $i \in N$, $f_i: [0,1] \rightarrow \mathbb{R}_{\leq 0}$), we will use the term ``bad cake''. We do not deal with general real-valued density functions in this work.

\paragraph*{Allocation:}
An \emph{allocation} $A \coloneqq (A_1,\dots,A_n)$ is given by $A_i = M_i \cup C_i$, where $(M_1, \ldots , M_n)$ is an $n$-partition of the set of indivisible items $M$, and $(C_1, \ldots , C_n)$ is an $n$-partition of the divisible resource $\C=[0,1]$. where $A_i$ is the \emph{bundle} allocated to the agent $i$ (note that $A_i$ can be empty). Given an allocation $A$, the \emph{utility} of agent $i \in N$ for the bundle $A_i$ is $v_i(A_i) \coloneqq v_i(M_i) + v_i(C_i)$, i.e., utility is additive across resource types.

\paragraph*{Perfect partition:}
For any $k \in \mathbb{N}$, a $k$-partition $C = (C_1, C_2, \ldots, C_k)$ of either cake or bad cake $\C$ is said to be perfect if each agent values all the pieces equally, i.e., for all agents $i \in N$ and for all pieces of the cake $j \in [k]$, $v_i(C_j) = \frac{v_i(\C)}{k}$. Note that a perfect allocation of a cake exists even when the agents' valuations are not normalized, since multiplicative scaling of agents' valuations preserves envy-freeness~\cite{A87splitting}. As in the work of Bei et al.~\cite{BLL+21fair}, we will assume the existence of a perfect allocation oracle in our algorithmic results.

\paragraph*{Generalized envy graph:}
The \emph{generalized envy graph} $\overline{G}_A$ of an allocation $A$ is a directed graph on the vertex set $N$, with a directed edge from agent $i$ to agent $k$ if $v_i(A_k) \ge v_i(A_i)$. If $v_i(A_k) = v_i(A_i)$, then we refer to the edge $(i,k)$ as an \emph{equality edge}, otherwise we call it an \emph{envy edge}. A \emph{generalized envy cycle} in this graph is a cycle $C$ that contains at least one envy edge.

\paragraph*{Top-trading generalized envy graph:}
The \emph{top-trading generalized envy graph} $\overline{T}_A$ of an allocation $A$ is a subgraph of $\overline{G}_A$, with a directed edge from agent $i$ to agent $k$ if $i \neq k$ and $v_i(A_k) = \max_{j \in N} v_i(A_j)$, i.e., $A_k$ is one of the most preferred bundles for agent $i$ in the allocation $A$. A generalized envy cycle in this graph is called a \emph{top-trading generalized envy cycle}.

\paragraph*{Envy-freeness for mixed resources (\EFM{}):}

We will now discuss the notion of envy-freeness for mixed resources (\EFM{}) that was formalized by Bei et al.~\cite{BLL+21fair} in the context of indivisible goods and divisible cake. Our definition extends their formulation to related settings where the indivisible part consists of chores and/or the divisible part is bad cake. The definition below is based on the following idea: Any agent who owns cake should not be envied, any agent who owns bad cake should not envy anyone else, and subject to these conditions, any pairwise envy should be \EF{1}. Formally, an allocation $A$ is said to be \emph{envy-free for mixed resources} (\EFM) if for any pair of agents $i, k \in N$, either $i$ does not envy $k$ (i.e., $v_i(A_i) \ge v_i(A_k)$), or all of the following hold: (a) $i$ does not have bad cake, i.e., $v_i(C_i) \ge 0$, (b) $k$ does not have cake, i.e., $v_i(C_k) \le 0$, and (c) the envy from $i$ to $k$ is bounded according to \EF{1}, i.e., $\exists \, j \in M_i \cup M_k$ such that $v_i(A_i \setminus \{j\}) \ge v_i(A_k \setminus \{j\})$.

\subsection{Background: Indivisible Goods and Cake}
\label{subsec:EF_Mixed_Resources_Introduction}

The algorithm of Bei et al.~\cite{BLL+21fair} gives an \EFM{} allocation for an instance with additive indivisible goods and cake. The algorithm initially finds an \EF{1} allocation of the indivisible goods using the envy-cycle elimination algorithm. It then allocates the cake in the following manner: In successive iterations, it tries to find an inclusion-wise maximal \emph{source addable set} of agents, to which cake is then allocated.

\begin{definition}[Source addable set]
Given a generalized envy graph, a non-empty set of agents $S \subseteq N$ is a \emph{source addable set} if (a) there is no envy edge from an agent in $N$ to an agent in $S$, and (b) there is no equality edge from an agent in $N \setminus S$ to an agent in $S$.
\end{definition}

Intuitively, to satisfy the \EFM{} property, an agent that is envied must not get any cake, and an equality edge $(i,j)$ implies that $i$'s value for the cake she gets must be at least her value for the cake that $j$ gets.

To find a maximal source addable set of agents, the algorithm first resolves all generalized envy cycles in the generalized envy graph $\overline{G}_A$. It then removes all agents that are reachable from an envied agent. Bei et al. show that the remaining agents form the unique maximal source addable set. If there are $k$ agents in this set, then the algorithm finds a perfect $k$-partition of the largest prefix of the cake (if $[a,1]$ is the remaining unallocated piece of cake, then a prefix of the cake is a piece $[a,x]$ of the cake where $a < x \le 1$) such that giving each agent in the set a piece of this partition does not introduce envy towards any agent in the set. This continues until all the cake is allocated.

\subsection{\EFM{}\ for Doubly Monotone Indivisible Items and Bad Cake}
\label{subsec:EFM_DM_Indivisible_Bad_Cake}

For an instance with doubly monotone indivisible items and bad cake, we give an algorithm to obtain an \EFM{} allocation (\Cref{alg:EFM_DM_Indivisible_Bad_Cake}). First, we run the doubly monotone algorithm (\Cref{alg:SEF1}) on the indivisible instance to obtain an \EF{1} allocation. We then extend it to an \EFM{} allocation by allocating the bad cake as follows: Our algorithm always allocates prefixes of the bad cake, hence the remaining cake is always an interval $[a, 1]$ for some $a \ge 0$. In each iteration, we first find an inclusion-wise maximal \emph{sink addable set} of agents, defined analogously to the source addable set introduced earlier.

\begin{definition}[Sink addable set]
    Given a generalized envy graph, a non-empty set of agents $S \subseteq N$ is a \emph{sink addable set} if (a) there is no envy edge from an agent in $S$ to an agent in $N$, and (b) there is no equality edge from an agent in $S$ to an agent in  $N \setminus S$.
\end{definition}

Since we will allocate bad cake to the agents in this set $S$, no agent in a sink addable set should envy another agent. Further, an equality edge $(i,j)$ implies that if $i$ is in the sink addable set $S$, $j$ must be in the set as well. We find a maximal sink addable set by first resolving all top-trading generalized envy cycles in the top-trading generalized envy graph $\overline{T}_A$, and then using the procedure in \Cref{lem:Uniq_Maximal_Sink_Addable}. The resolution of top-trading generalized envy cycles does not affect the \EFM{} property, because of the same reasons as in the top-trading envy-cycle elimination algorithm (\Cref{sec:SEF1}).

Once we find the maximal sink addable set $S$ to which we can allocate bad cake, we need to quantify the amount of bad cake that can be allocated to the agents in $S$ while still preserving the \EFM{} property. We find the largest loss in utility $\delta_i$ that an agent $i \in S$ (who is to be given bad cake) can tolerate before she starts to envy another agent $j \not\in S$ (who is not allocated any bad cake), i.e., \[
\delta_i = \min_{j \in N\setminus S} v_i(A_i) - v_i(A_j) \text{ for all }i \in S
.\] Note that $\delta_i > 0$ since there are no envy or equality edges from $S$ to  $N \setminus S$. We then find the smallest prefix $[a, x_{i^*}]$ of the cake, and a perfect $|S|$-partition of this prefix, so that if each part is allocated to an agent in $S$, then the utility of each agent decreases by at most $\delta_i$, and for a particular agent $i^*$, her utility goes down by exactly $\delta_{i^*}$. By definition of $\delta_{i^*}$, a new equality edge arises in the generalized envy graph $\overline{G}_A$ from agent $i^* \in S$ to some agent in $N\setminus S$. Once we allocate $[a,x_{i^*}]$ perfectly to all agents in $S$, the allocation still remains \EFM{} (\Cref{lem:Partial_Allocation_EFM}), and we only have $[x_{i^*}, 1]$ of the cake left to allocate. We will establish in \Cref{thm:EFM_DM_Bad_Cake_Termination} that the algorithm terminates with a polynomial number of such iterations.

\begin{algorithm}[t]
\DontPrintSemicolon
\KwIn{An instance $\langle N, M, \C, \V, \F \rangle$ with doubly monotone indivisible items $M$, and a divisible bad cake $\C$}
\KwOut{An allocation $A$}
\BlankLine
Run the doubly monotone algorithm to obtain an \EF{1} allocation $A = ( A_1, A_2, \ldots, A_n)$ of $M$\;
\tcp{Bad cake allocation phase}
\While{there is still unallocated cake $\C= [a,1]$} {
    $\overline{T}_A = $ top-trading generalized envy graph of $A$\;
    \While{ there is a top-trading generalized envy cycle $C$ in  $\overline{T}_A$} {
        $A \leftarrow A^C$ \Comment*[r]{\footnotesize{This ensures the existence of a sink addable set}}
    }
    $S = $ maximal sink addable set for $A$ \Comment*[r]{\footnotesize{Using \Cref{lem:Uniq_Maximal_Sink_Addable,lem:No_TTC_Implies_Sink_Addable}}}
    \If{$S=N$} {
    Find an \EF{} allocation $(C_1, C_2, \ldots, C_n)$ of $\C$\;
    $A_i = A_i \cup C_i$ for all  $i \in N$\;
    $\C \leftarrow \emptyset$ \;
    }
    \Else{
        $ \delta_i = \min_{j \in N\setminus S} v_i(A_i) - v_i(A_j)$ for all $i \in S$\;
        \If{$v_i(\C) \ge - |S| \cdot \delta_i$ for all $i \in S$}{
            $C' \leftarrow \C$\;
             $\C \leftarrow \emptyset$\;
        }
        \Else{
            $x_i = \sup \left\{ x \mid v_i([a,x]) \ge - |S| \cdot \delta_i \right\} $ for all $i \in S$\;
            $i^* = \arg \min_{i \in S} x_i$\;
            $C' \leftarrow [a, x_{i^*}]$\;
            $\C \leftarrow [x_{i^*}, 1]$ \;
        }
        Obtain a perfect partition $(C_1, C_2, \ldots, C_{|S|})$ of $C'$\;
        $A_i \leftarrow A_i \cup C_i$ for all  $i \in S$\;
    }
}
\KwRet $A$
\caption{Algorithm for \EFM{} with doubly monotone indivisible items and bad cake}
\label{alg:EFM_DM_Indivisible_Bad_Cake}
\end{algorithm}

We first show that the maximal sink addable set is unique if it exists (Bei et al.~\cite{BLL+21fair} show a similar result for source addable sets). All missing proofs can be found in the appendix.

\begin{restatable}{lemma}{UniqMaximalSinkAddable}
\label{lem:Uniq_Maximal_Sink_Addable}
Given a partial allocation $A$, the maximal sink addable set (if it exists) is unique, and can be found in polynomial time.
\end{restatable}

We now show that once we resolve all top-trading generalized envy cycles, the generalized envy graph $\overline{G}_A$ contains a sink addable set (and thus a maximal sink addable set).

\begin{restatable}{lemma}{NoTTCImpliesSinkAddable}
\label{lem:No_TTC_Implies_Sink_Addable}
If the top-trading generalized envy graph $\overline{T}_A$ does not contain any generalized envy cycles, then the generalized envy graph $\overline{G}_A$ has a sink addable set.
\end{restatable}

Once we run the top-trading generalized envy cycle elimination procedure, we are thus guaranteed the existence of a sink addable set. Then, we move to the bad cake allocation procedure. The agents in the set $S$ are then perfectly allocated a small amount of bad cake while preserving the \EFM{} property. We now show that the partial allocation remains \EFM{} throughout the algorithm.

\begin{restatable}{lemma}{PartialAllocationEFM}
\label{lem:Partial_Allocation_EFM}
At each step of the bad cake allocation phase, the partial allocation in \Cref{alg:EFM_DM_Indivisible_Bad_Cake} satisfies \EFM{}.
\end{restatable}

Finally, we will show that the algorithm terminates, assuming the existence of a perfect partition oracle.

\begin{restatable}{theorem}{EFMDMBadCakeTermination}
\label{thm:EFM_DM_Bad_Cake_Termination}
\Cref{alg:EFM_DM_Indivisible_Bad_Cake} terminates after $\O(n^3)$ rounds of the while-loop and returns an \EFM{} allocation.
\end{restatable}

\subsection{Special Case Results with Indivisible Chores and Divisible Cake}
\label{subsec:EFM_Special_Cases_Chores_and_Cake}

While the approach of first allocating the indivisible resources followed by assigning the divisible resource works well for instances with indivisible goods and cake~\cite{BLL+21fair}, and for instances with doubly monotone indivisible items and bad cake (\Cref{thm:EFM_DM_Bad_Cake_Termination}), extending this approach to an instance with indivisible chores and cake is challenging for the following reason: Suppose, in such an instance, we initially allocate the indivisible chores to satisfy \EF{1} using the top-trading envy-cycle elimination algorithm. Then we might not be able to proceed with cake allocation, since the algorithm does not guarantee us the existence of a source in the generalized envy graph. In an effort to remedy this, we introduce the component-wise matching algorithm (presented in \Cref{subsec:Componentwise_Matching_Algo}) to obtain an \EF{1} allocation of additive indivisible chores that does not have any generalized envy cycles. This algorithm ensures that the allocation at the end of indivisible chores stage is generalized envy-cycle free. However, adding even a small amount of cake might once again make the generalized envy graph sourceless, and it is unclear how to proceed at this stage.

Nevertheless, for special cases of this problem, we can prove the existence of an \EFM{} allocation. Restricted to additive valuations of the indivisible chores, we show methods of obtaining an \EFM{} allocation in two cases: 1) when the agents have \emph{identical rankings} of the items (formalized in \Cref{subsec:EFM_Special_Cases_Chores_and_Cake_Appendix}), and 2) when the number of items does not exceed the number of agents by more than one, i.e., $m \le n + 1$. At a high level, the reason we are able to circumvent the aforementioned challenge in these two cases is because the particular \EF{1} allocation we obtain for indivisible chores in these cases has the property that we can resolve \emph{any} generalized envy cycle that arises during the cake allocation stage. This freedom allows us to execute the algorithm of Bei et al. directly on these instances.

\begin{restatable}{theorem}{EFMChoresCakeIdenticalRankings}
\label{thm:EFM_Chores_Cake_Identical_Rankings}
For a mixed instance with additive indivisible chores with identical rankings and cake, an \EFM{} allocation exists.
\end{restatable}

\begin{restatable}{theorem}{EFMChoresCakeSmallNoofItems}
\label{thm:EFM_Chores_Cake_Small_No_of_Items}
For a mixed instance with $n$ agents, $m$ additive indivisible chores and cake where $m \le n+1$, an \EFM{} allocation exists.
\end{restatable}

The proofs of the above two theorems can be found in \Cref{subsec:EFM_Special_Cases_Chores_and_Cake_Appendix}.

\bibliography{References}

\newpage

\appendix

\section{Appendix}

\subsection{Proof of Theorem~\ref{thm:EF_NP-complete_Chores_Binary}}
\label{subsec:Proof_NPC_Chores_Binary}

\NPhardness*

\begin{proof}
Membership in \NP{} follows from the fact that given an allocation, checking whether it is envy-free can be done in polynomial time.

To show \NPH{}ness, we will show a reduction from \SetSplitting{} which is known to be \NPC{}~\cite{GJ79computers} and asks the following question: Given a universe $U$ and a family $\F$ of subsets of $U$, does there exist a partition of $U$ into two sets $U_1,U_2$ such that each member of $\F$ is \emph{split} by this partition, i.e., no member of $\F$ is completely contained in either $U_1$ or $U_2$?

\emph{Construction of the reduced instance}:
Let $q \coloneqq |U|$ and $r \coloneqq |\F|$ denote the cardinality of the universe $U$ and the set family $\F$, respectively. Let $r' \coloneqq \max\{q,r\}$. We will find it convenient to refer to the universe as a set of `vertices', the members of the set family $\F$ as a set of `hyperedges', and the membership in $U_1$ or $U_2$ as each vertex being `colored' $1$ or $2$.

We will construct a fair division instance with $m = r'+q$ chores and $n = r' + 2$ agents. The set of chores consists of $r'$ \emph{dummy} chores $D_1,\dots,D_{r'}$ and $q$ \emph{vertex} chores $V_1,\dots,V_q$. The set of agents consists of $r'$ \emph{edge} agents $e_1,\dots,e_{r'}$, and two \emph{color} agents $c_1,c_2$. When $r' = r$ (i.e., $r \geq q$), each edge agent should be interpreted as corresponding to a hyperedge, and otherwise if $r < r'$, then we will interpret the first $r$ edge agents $e_1,\dots,e_{r}$ as corresponding to the hyperedges while each of the remaining edge agents $e_{r+1},\dots,e_{r'}$ will be considered as an ``imaginary hyperedge'' that is adjacent to the entire set of vertices (and therefore does not impose any additional constraints on the coloring problem).

\emph{Preferences}: The valuations of the agents are specified as follows: Each dummy chore is valued at $-1$ by all (edge and color) agents. Each vertex chore $V_j$ is valued at $-1$ by those edge agents $e_i$ whose corresponding hyperedge $E_i \in E$ is adjacent to the vertex $v_j \in V$, and at $0$ by all other edge agents. The color agents value all vertex chores at $0$. This completes the construction of the reduced instance. We will now argue the equivalence of the solutions.

($\Rightarrow$) Suppose there exists a partition of the universe $U$ that splits all member of $\F$ (equivalently, a feasible $2$-coloring of the corresponding hypergraph such that each hyperedge sees both colors). Then, an envy-free allocation can be constructed as follows: The $r'$ dummy chores are evenly distributed among the $r'$ edge agents. In addition, if the vertex $v_j$ is assigned the color $\ell \in \{1,2\}$, then the vertex chore $V_j$ is assigned to the color agent $c_\ell$.

The aforementioned allocation is feasible as it assigns each item to exactly one agent. Furthermore, it is also envy-free for the following reason: Each color agent only receives vertex chores and has utility $0$, and therefore it does not envy anyone else. The utility of each edge agent $e_i$ is $-1$ because of the dummy chore assigned to it. However, $e_i$ does not envy any other edge agent $e_\ell$ since the latter is also assigned a dummy chore. Furthermore, $e_i$ also does not envy any of the color agents since, by the coloring condition, each of them receives at least one chore that is valued at $-1$ by $e_i$.

($\Leftarrow$) Now suppose that there exists an envy-free allocation, say $A$. Then, it must be that none of the color agents receive a dummy chore. This is because assigning a dummy chore to a color agent $c_\ell$ would give it a utility of $-1$, and in order to compensate for the envy, it would be necessary to assign \emph{every} other agent at least one chore that $c_\ell$ values at $-1$. This, however, is impossible since the number of agents other than $c_\ell$ is $r'+2$, which strictly exceeds the number of chores that $c_\ell$ values at $-1$, namely $r'$. Thus, all dummy chores must be allocated among the edge agents.

We will now show that no edge agent receives more than one dummy chore under the allocation $A$. Suppose, for contradiction, that the edge agent $e_i$ is assigned two or more dummy chores. Then, due to envy-freeness, every other agent must get at least two chores that $e_i$ values at $-1$. There are $r'+2$ agents in total excluding $e_i$, which necessitates that there must be at least $2r'+4$ chores valued at $-1$ by $e_i$. However, the actual number of chores valued at $-1$ by $e_i$ that are available for allocation is at most $(r'-2)+q$, which, by the choice of $r'$, is strictly less than $2r'+4$, leading to a contradiction. Thus, each edge agent receives at most one dummy chore, and since the number of edge agents equals that of dummy chores, we get that the $r'$ dummy chores are, in fact, evenly distributed among the $r'$ edge agents.

Since each edge agent $e_i$ receives a dummy chore, in order to compensate for the envy the allocation $A$ must assign each color agent at least one vertex chore that $e_i$ values at $-1$. The desired coloring for the hypergraph (equivalently, the desired partition of the universe $U$) can now be naturally inferred; in particular, the elements of $U$ corresponding to the vertex chores whose assignment is not forced by the aforementioned remark can be put in an arbitrary partition. This completes the proof of \Cref{thm:EF_NP-complete_Chores_Binary}.
\end{proof}

\subsection{Proof of Lemma~\ref{lem:SEF1_Goods}}
\label{subsec:Proof_SEF1_Goods}

\SEFGoods*
\begin{proof}
Clearly the empty allocation at the beginning is \EF{1}. Suppose before time step $t$, our allocation $A$ is \EF{1} (i.e., any envy from agent $i$ to agent $j$ can be eliminated by removing an item from $A_j$). Denote the allocation after time step $t$ by $A'$. We will argue that $A'$ is \EF{1}, and any envy from agent $i$ to agent $j$ can be eliminated by removing an item from $A_j'$. At every time step, either a good is allocated or an envy cycle is resolved.

Suppose at time step $t$, we allocate a new item (Lines 5-6). Note that the graph $G_A$ is acyclic at this stage. This is because it holds trivially the first time an item is allocated, and in every subsequent execution of the while-loop, we eliminate all envy cycles present (Lines 7-8) before we begin allocating the next ite. Thus, the subgraph $G_A^g$ is acyclic as well, where $G_A^g$ is the graph $G_A$ restricted to the agents for whom $g$ is a good.

Then after time $t$, our allocation $A'$ will be $A_k' = A_k \cup \left\{ g \right\} $, and $A'_j = A_j$ for all  $j \neq k$, where $k$ is a source in $G_A^g$. Pick two agents $i$ and $j$ such that $i$ envies $j$ in $A'$. If $i$ did not envy $j$ in $A$, then clearly $j$ must be the agent who received the good $g$ (i.e., $j = k$) and $i \in V^g$. In this case, removing $g$ from $A_k$ removes $i$'s envy as well. Suppose $i$ envied $j$ in $A$ as well, and the envy was eliminated by removing $g'$ from $A_j$. If $j=k$ then $i \not\in V^g$ since $k$ was a source in $G_A^g$.  Then removing $g'$ eliminates the envy in $A'$ as well, since $v_i(A_k \cup \{g\} \setminus \{g'\}) \le v_i(A_k \setminus \{g'\}) $. If $j \neq k$, then since $j$'s bundle remains the same and $v_i(A_i') \ge v_i(A_i)$, the envy can again be eliminated by removing $g'$ from $A_j'$.

Suppose at time $t$ we resolve an envy cycle (Lines 7-8). Let $A$ be the allocation before time $t$, $C$ be the cycle along which the swap happens, and $A' = A^C$ the allocation obtained by swapping backwards along the circle. Pick two agents $i$ and $j$ such that $i$ envies $j$ in $A'$. Let $i'$ and $j'$ be the agents such that $A_i' = A_{i'}$ and $A_j' = A_{j'}$. Since $v_i(A_i') \ge v_i(A_i)$, $i$ envied $j'$ in the allocation $A$ before the swap. Suppose this envy was eliminated by removing $g'$ from $A_{j'}$. Then $v_i(A_i') \ge v_i(A_i) \ge v_i(A_{j'} \setminus \{g'\})$, and thus removing $g'$ from $A_j'$ eliminates the envy in $A'$.

To show that the algorithm terminates in polynomial time, we show that we resolve envy cycles at most a polynomial number of times for each item. Consider a single while-loop for an item, where a cycle swap occurs on the cycle $C$. Since the bundles remain unbroken, all agents outside the cycle have the same outdegree in $G_{A'}$ as in $G_{A}$. An agent $i$ inside the cycle has strictly lesser outdegree in $G_{A'}$ compared to $G_{A}$, since the $(i,i^+)$ edge in $G_A$ does not translate into a $(i,i)$ edge in $G_{A'}$ (since $i$ gets $i^+$'s bundle). Thus the number of envy edges goes down by at least $|C|$ during each cycle swap, and the while-loop terminates in polynomial time.
\end{proof}

\subsection{Proof of Lemma~\ref{lem:Uniq_Maximal_Sink_Addable}}
\label{subsec:Proof_Uniq_Maximal_Sink_Addable}

\UniqMaximalSinkAddable*
\begin{proof}
Suppose there were two distinct maximal sink addable sets $S_1$ and $S_2$. Then we show that $S_1 \cup S_2$ is also a sink addable set, contradicting their maximality. The existence of an envy edge from an agent in $S_1 \cup S_2$ to an agent in $N$ contradicts either $S_1$ or $S_2$ being a sink addable set. Similarly, an equality edge from an agent in $S_1 \cup S_2$ to an agent in $N \setminus S_1 \cup S_2$ contradicts either $S_1$ or $S_2$ being a sink addable set. Thus $S_1 \cup S_2$ is a sink addable set, a contradiction.

To find a maximal sink addable set, say that an agent $i$ is an \emph{envious} agent if there is an envy edge $(i,j)$ in the generalized envy graph. Note that by definition a sink addable set cannot contain an envious agent. Let $T$ be the set of all agents that have a path to an envious agent (i.e., if agent $r$ is in $T$, there there exists a path from $r$ to an envious agent $i$ in the generalized envy graph along envy and equality edges). Let $S = N \setminus T$ be the complementary set of agents. We claim that $S$ is a maximal sink addable set. Clearly, no agent outside $S$ can be in any sink addable set. To see this, consider any agent $r \not \in S$, and let $r$ have a path to an envious agent $i$. By the properties of sink addable sets, if $r$ is in $S$, then so must all the agents in the path including $i$, but this contradicts the property that a sink addable set cannot contain an envious agent. Now consider an agent $r$ in $S$, and note that $r$ does not have a path to an envious agent. Since $r$ is not envious, it does not have an envy edge to any other agent. Further, since all agents outside $S$ have a path to an envious agent, $r$ cannot have an (equality or envy) edge to an outside agent. Hence, the set of agents so obtained must be a maximal sink addable set.
\end{proof}

\subsection{Proof of Lemma~\ref{lem:No_TTC_Implies_Sink_Addable}}
\label{subsec:Proof_No_TTC_Implies_Sink_Addable}

\NoTTCImpliesSinkAddable*
\begin{proof}
Suppose the top-trading generalized envy graph $\overline{T}_A$ does not contain any generalized envy cycles. Then each strongly connected component $C_i$ of the graph $\overline{T}_A$ contains only equality edges inside it. Let $C_1$ be a leaf component obtained by Tarjan's algorithm to find strongly connected components (see Section 22.5 of~\cite{CLRS}). We claim that $C_1$ is a sink addable set in the generalized envy graph $\overline{G}_A$.

Suppose there was an envy edge from an agent $i$ in $C_1$. Since the top-trading envy graph points to an agent's favorite bundle, there would be an envy edge from $i$ in the graph  $\overline{T}_A$ as well. Thus $i$ would not be part of a leaf component of $\overline{T}_A$, contradicting that  $i \in C_1$. Thus all agents in $C_1$ only have equality edges in $\overline{G}_A$.

Suppose now that there was an equality edge from an agent $i \in C_1$ to an agent $j \in N \setminus C_1$. Since $i$ does not envy any other agent, the edge  $(i, j)$ would be present in  $\overline{T}_A$ as well, contradicting that $C_1$ is a leaf component of $\overline{T}_A$. Thus  $C_1$ is a sink addable set, and thus $\overline{G}_A$ contains a maximal sink addable set as well.
\end{proof}

\subsection{Proof of Lemma~\ref{lem:Partial_Allocation_EFM}}
\label{subsec:Proof_Partial_Allocation_EFM}

\PartialAllocationEFM*

\begin{proof}
The allocation of indivisible items at the start of the algorithm is \EF{1} and consequently \EFM{} as well. In the generalized top-trading envy graph, suppose we resolve a cycle $C$. Then the allocation remains \EFM{} for agents outside the cycle since the bundles are unbroken. Since all agents in the cycle receive their highest valued item, they do not envy any other agent and thus satisfy \EFM{} as well.

In the bad cake allocation stage, if $N$ is the maximal sink addable set, then there are no envy edges inside the graph $\overline{G}_A$ and the initial allocation is envy-free. The allocation remains envy-free on adding an \EF{} allocation of the remaining cake to their bundles, and thus the final allocation is \EFM{} as well.

If the maximal sink addable set $S \subset N$ is a strict subset, then the amount of cake we allocate is chosen such that the \EFM{} property is satisfied. Since every agent in $S$ is given bad cake, the envy from agents in  $N \setminus S$ remains \EFM{}. By definition of sink addable set, none of the agents in $S$ envy anyone before the bad cake allocation. Since we obtain a perfect allocation, none of the agents in $S$ envy each other after the bad cake allocation as well. Note that the value of the bad cake allocated to each agent in this round is bounded below by $ \delta_i = \min_{j \in N\setminus S} v_i(A_i) - v_i(A_j)$ for all $i \in S$. Thus no agent in $S$ envies an agent in  $N \setminus S$ in the final allocation as well by choice of $\delta_i$, and the partial allocation remains \EFM{} throughout the algorithm.
\end{proof}

\subsection{Proof of Theorem~\ref{thm:EFM_DM_Bad_Cake_Termination}}
\label{subsec:Proof_EFM_DM_Bad_Cake_Termination}

We will break down the running time analysis into two parts: (1) The number of rounds where the algorithm resolves a top-trading generalized envy cycle, and (2) between any such consecutive rounds, the number of times when the algorithm assigns bad cake to agents in a maximal sink addable set.

Let us start with the first part. Note that assigning bad cake to a maximal sink addable set never creates new envy edges, and resolving a top-trading generalized envy cycle reduces the number of envy edges by at least one. Therefore, following the allocation of the indivisible items, the number of envy edges is a non-increasing function of time. This means that there can be at most $\O(n^2)$ rounds where a top-trading generalized envy cycle is resolved by the algorithm.

Let us now consider the second part. We will argue that between any consecutive rounds where the algorithm resolves a top-trading generalized envy cycle, there can be at most $n$ steps where the algorithm allocates bad cake. Observe that if there are no envy edges in the graph, then the maximal sink addable set is the entire set of agents (i.e., $S = N$), and the algorithm immediately terminates by allocating the entire remaining cake. Otherwise, after each round of adding bad cake, at least one new equality edge is created from $S$ to  $N \setminus S$. If this creates a new top-trading generalized envy cycle, then the number of envy edges strictly reduces in the next round. Else, the size of the maximal sink addable set strictly decreases in the next round, implying that there can be at most $n$ rounds of adding bad cake before the number of envy edges strictly decreases.

Overall, we obtain that the algorithm terminates in $\O(n^3)$ rounds.

\EFMDMBadCakeTermination*
\begin{proof}
By \Cref{lem:Partial_Allocation_EFM}, the allocation returned by the algorithm is \EFM{}. So, it suffices to show that the algorithm executes $\O(n^3)$ iterations of the while-loop.

First, note that there can be at most $\O(n^2)$ rounds where the algorithm resolves a top-trading generalized envy cycle. This follows from the following two observations: (a) The number of envy edges never increases during the algorithm, and (b) the number of envy edges strictly decreases by at least one whenever a top-trading generalized envy cycle is resolved. Observation (b) is straightforward. To see why (a) holds, it suffices to argue that adding bad cake does not introduce any new envy edges. Indeed, since we are adding bad cake, none of the agents in the set $N \setminus S$ (i.e., the agents outside the maximal sink addable set) can develop new envy edges to an agent in $S$ or $N \setminus S$. For each agent in $S$, since the amount of cake added is bounded below by $ \delta_i = \min_{j \in N\setminus S} v_i(A_i) - v_i(A_j)$ for all $i \in S$, none of the agents in $S$ have new envy edges to an agent in  $N \setminus S$. Furthermore, since the allocation is perfect, agents in $S$ do not end up envying one another. Thus, no new envy edges are created due to the allocation of the bad cake, and therefore we have at most $\O(n^2)$ rounds of top-trading generalized envy-cycle elimination.

Next, we will argue that between any consecutive cycle-elimination rounds, there can be at most $n$ steps where the algorithm assigns bad cake to a maximal sink addable set. Note that if at any stage there are no envy edges in the generalized envy graph, then the maximal sink addable set is the entire set of agents (i.e., $S = N$), and the algorithm terminates immediately after assigning the entire remaining bad cake. So, let us assume for the remainder of the proof that there is always at least one envy edge.

If, for all agents $i \in S$, a perfect allocation of the remaining cake $\C = [a, 1]$ preserves \EFM{}, then the algorithm terminates in the next step. Else, we mark the point $x_i$ on the cake for each agent such that after perfectly allocating $[a, x_i]$, they have a new equality edge to an agent in  $N \setminus S$. By choice of $i^*$ as the agent with minimum value of $x_i$, the agent  $i^*$ has a new equality edge to an agent $j$ in  $N \setminus S$ after this round. If this edge creates a new top-trading generalized envy cycle, then the number of envy edges strictly reduces in the next round. Else, the size of the maximal sink addable set decreases since $i^*$ now has a path to an envious agent and must therefore be excluded from the maximal sink addable set (additionally, no new agents are added to the sink addable set). Thus, bad cake allocation can occur for at most $n$ consecutive rounds before the number of envy edges strictly decreases, leading to the desired $\O(n^3)$ number of rounds. \end{proof}

\subsection{Special Case Results for \EFM{} with Indivisible Chores and Divisible Cake}
\label{subsec:EFM_Special_Cases_Chores_and_Cake_Appendix}

We first discuss the case of additive indivisible chores with identical rankings and cake. By identical rankings, we mean that all the agents have the same preference order on the indivisible chores, and we can order the chores as $ c_1, c_2, \ldots, c_m$ such that $v_i(c_1) \ge v_i(c_2) \ge \ldots \ge v_i(c_m)$ for all agents $i \in N$. We assume for simplicity that the number of chores is a multiple of the number of agents. If not, we add an appropriate number of virtual chores that are valued at 0 by every agent, and remove them at the end of the algorithm. Note that this does not affect the \EF{1} or the \EFM{} property. In this setting, the round-robin algorithm satisfies two desirable properties:
\begin{itemize}
    \item Let $(B_1, B_2, \ldots, B_n)$ be a partition of the chores given by $B_i = \{ c_{j} \mid j \equiv i \pmod{n} \}$. For each of the $n!$ possible orderings of the agents, under lexicographic tiebreaking of the preferable chores, the round-robin algorithm allocates the same bundle $B_i$ to the $i^{th}$ agent in the ordering, and
    \item Resolving \emph{any} generalized envy cycle in the allocation obtained from a round-robin instance does not violate \EFM{}.
\end{itemize}
By lexicographic tiebreaking, we mean that if an agent has many chores of the same value to choose in round-robin, they choose the chore with the lexicographically smallest index.

\begin{restatable}{lemma}{IdenticalRankingsRoundRobin}
\label{lem:Identical_Rankings_Round_Robin}
For an additive indivisible chores instance with identical rankings, every ordering of the agents for round-robin allocates the bundle $B_i = \{ c_{j} \mid j \equiv i \pmod{n} \}$ to the $i^{th}$ agent in the ordering when tiebreaking happens lexicographically.
\end{restatable}

\begin{proof}
Order the chores as $ c_1 , c_2, \ldots, c_m$ in non-increasing order of their value. Suppose that in the execution of the round-robin algorithm, if an agent has many chores that it has the same value for, it chooses the lexicographically smallest chore. We claim that the bundles remain the same regardless of the ordering of the agents.

Suppose the agents were ordered as $\pi(1), \pi(2), \ldots, \pi(n)$ for round-robin. In the first round of the algorithm, we claim that agent $\pi(i)$ chooses the chore $c_i$ during the execution. Indeed, since all the agents have the same rankings on the chores, agent $\pi(1)$ chooses $c_1$ in the first round (even if there were other chores with the same value, $c_1$ is lexicographically the smallest). Inductively, once agents $\{ \pi(1), \pi(2), \ldots, \pi(i) \}$ have chosen $\{c_1, c_2, \ldots, c_i\}$, agent $\pi(i+1)$ weakly prefers $c_{i+1}$ over any other chore, and adds it to their bundle because it is the lexicographically smallest chore present. Thus $\{c_1, c_2, \ldots, c_n \}$ are allocated in the first round. By a straightforward induction on the number of rounds, we see that the final allocation is $A_{\pi(i)} = \{ c_{j} \mid j \equiv i \pmod{n} \}$.
\end{proof}

Note that for every position $j$ and every agent $i$, there is an ordering of the agents such that agent $i$ is in the  $j^{th}$ position during the round-robin algorithm (if $i = j$ consider the identity permutation, else consider the permutation $(i\ j)$). Since the round-robin algorithm always returns an \EF{1} allocation, this implies the strong property that the partition $(B_1, B_2, \ldots, B_n)$ satisfies \EF{1} for agent $i$, regardless of which bundle is allocated to that agent. Since the agent $i$ was chosen arbitrarily at the beginning, this gives us the following lemma:

\begin{restatable}{lemma}{Identical_Rankings_Very_Strong_EF}
\label{lem:Identical_Rankings_Very_Strong_EF1}
For an additive indivisible chores instance with identical rankings, the allocation $A_{\pi(i)} = B_i$ is an \EF{1} allocation for any ordering $\pi$ of the agents, where $B_i = \{ c_{j} \mid j \equiv i \pmod{n} \}$.
\end{restatable}

Recall from \Cref{subsec:Monotone_Chores} that the reason we had to restrict ourselves to resolving top-trading envy cycles when searching for an \EF{1} allocation of monotone indivisible chores was because resolving an arbitrary envy cycle could upset \EF{1} (\Cref{eg:Envy_Graph_Example_Intro}), where an agent might obtain a bundle of higher value during the cycle swap, but the newly acquired bundle might consist of chores with low absolute value. By contrast, we can resolve \emph{any} generalized envy cycle in the case of identical rankings, since the allocation is \EF{1} regardless of which agent obtains which bundle. Note that this property continues to hold even when we add cake using Bei et al.'s algorithm, since any agent with cake is never envied throughout the algorithm, and any envy present can be eliminated by the removal of one good (in fact, the same good) regardless of the identity of the agent holding a bundle (the valuation is weakly better for bundles with some portion of cake present). Thus, we have the following theorem:

\EFMChoresCakeIdenticalRankings*

Using this result, we also show the existence of an \EFM{} allocation when $n-1$ of the $n$ agents have identical rankings. Say the agents are $\{1, 2, \ldots, n\}$, and the agent $n$ does not have a ranking identical to the others. Run the round-robin algorithm with the first $n-1$ agents and one virtual copy of one of the identical agents, say agent $1$. At the end of the round-robin algorithm, allow agent $n$ to pick their favorite bundle and arbitrarily allocate the remaining bundles between the other agents. Note that since agent $n$ does not envy anybody at the beginning of cake allocation, and since the cake allocation stage does not create any new envy edges, the final allocation will be \EFM{} for agent $n$. For all other agents, the allocation will be \EF{1} at the beginning of cake allocation (\Cref{lem:Identical_Rankings_Very_Strong_EF1}). Since resolving any generalized envy cycle still preserves \EFM{} (\Cref{thm:EFM_Chores_Cake_Identical_Rankings}), the final allocation will be \EFM{} for the agents  $\{1, 2, \ldots, n-1 \}$ as well. Thus we get the following corollary:

\begin{restatable}{corollary}{EFM_Chores_Cake_Identical_Except_One}
\label{cor:EFM_Chores_Cake_Identical_Except_One}
For a mixed instance with additive indivisible chores with identical rankings for $n-1$ agents and cake, an \EFM{} allocation exists. In particular, for two agents, an \EFM{} allocation always exists in this setting.
\end{restatable}

This property of being able to resolve any generalized envy cycle can also be obtained when the number of indivisible chores is at most one higher than the number of agents, i.e., $m \le n+1$, which allows us to obtain an \EFM\ allocation in this setting.

\EFMChoresCakeSmallNoofItems*

\begin{proof}
We first show that we can resolve any generalized envy cycle when the number of indivisible chores is at most one higher than the number of agents, i.e., $m \le n+1$. When $m \le n$, this is trivial since we can initially allocate any single chore to each agent. In this case, subsequent resolution of any generalized envy cycles while allocating cake using the algorithm of Bei et al. preserves the \EFM{} property, since any envied bundle is always devoid of cake, while dropping the single chore from an envious agent's bundle makes his valuation for his own bundle non-negative, thereby satisfying \EFM{}. Suppose $m = n + 1$. On execution of the round-robin algorithm on the indivisible chores, we obtain an allocation where one agent has two chores $\{c_1, c_2\}$ and all the other agents have a single chore. In the cake allocation stage, since the amount of cake to be allocated in each round is carefully chosen so as to preserve \EFM{}, we only need to show that \EFM{} is preserved when resolving any generalized envy cycle that might arise as a result of cake allocation.

Clearly any cycle resolution preserves \EFM{} for an agent receiving a bundle with a single chore because of the same arguments as in the case when $m \le n$. Suppose an agent $i$ receives the bundle $\{c_1, c_2, C \}$ during cycle resolution, where $C$ is some piece of cake. We claim that dropping $c_2$ preserves \EFM{}. Since \EFM{} is trivially preserved for all agents that $i$ does not envy, suppose $i$ envies an agent $j$. Since $j$'s current bundle was envied by $i$ before this round as well, $j$'s current bundle cannot contain any cake, and only contains a single chore.  Suppose agent $i$ had the chore $c$ allocated to her during the round-robin stage. Since cake is non-negatively valued by all agents, and since cycle resolution does not decrease any agent's valuation for their bundle, $v_i(c_1) + v_i(c_2) + v_i(C) \ge v_i(c)$. Since she chose $c$ over $ c_2$ in round-robin, $v_i(c) \ge v_i(c_2)$. Putting these inequalities together, we get that agent $i$ has non-negative valuation for her new bundle on dropping $c_2$. Since $j$'s bundle only contains an indivisible chore, cycle resolution preserves \EFM{} for agent $i$ as well. Thus, we get that:
\end{proof}

\subsection{Component-wise Matching Algorithm}
\label{subsec:Componentwise_Matching_Algo}

While Algorithm~\ref{alg:SEF1} returns an \EF{1} allocation, there might be unresolved envy cycles even in the final allocation. A generalized envy cycle in the generalized envy graph $\overline{G}_A$ implies an obvious Pareto improvement that we may not be able to perform because it might destroy the \EF{1} property. Recall that a generalized envy cycle is a cycle in $\overline{G}_A$ that contains at least one envy edge. For a monotone instance with additive valuations and indivisible chores, we give an algorithm that returns an \EF{1} allocation that is generalized envy cycle free.

The problem with obtaining an \EF{1} allocation without envy cycles for chores was the fact that in Algorithm~\ref{alg:SEF1}, one could not remove any envy cycle of their choice. For the case of additive chores, we present a modified version of the round robin algorithm where we maintain the invariant that after each round of the algorithm, the allocation is generalized envy cycle free. In each round, we move from the sink strongly connected components (referred to henceforth as components) all the way up to the source, and perform a maximum weight perfect matching in the bipartite graph $H_j = (S_j \cup M, E)$, with the vertices of the component $S_j$ on one side and the remaining unallocated items on the other. The weight of an edge from agent $i$ to chore $c$ is its value $v_i(c)$ for the chore. We show that this returns an \EF{1} allocation without generalized envy cycles.

At the start of each round, we first find a topological sorting of the components (see Section 22.5 of \cite{CLRS}). Let $\text{ComponentToposort}(\cdot)$ be the subroutine that takes in a directed graph $G$ and returns $S = (S_1, S_2, S_3, \ldots S_{\ell})$, the components of $G$ in topological order. There are no new cycles created within a component after the round since that would contradict maximality of the matching, and there are no new cycles between components either because we allocated chores in reverse topological order. Thus the allocation is generalized envy cycle free, and properties of the round robin algorithm assure us that this allocation is \EF{1}.

To make analysis easier, we assume that the number of items is a multiple of the number of agents. If not, we add virtual items that are valued at $0$ by every agent, and remove them at the end of the algorithm. Note that this preserves the \EF{1} property.

\begin{algorithm}[t]
\DontPrintSemicolon
\KwIn{$\langle A, N, M, \V \rangle$ where $A = (A_1, A_2, \ldots , A_n)$ is a partial allocation, $N$ a set of agents, $M$ a set of unallocated chores, and $\V$ a valuation function}
\KwOut{An allocation $A$}

\If{$M = \emptyset$} {
    \KwRet $A$
}
\Else{
    $S \leftarrow \text{ComponentToposort}(\overline{G}_A)$ \Comment*[r]{\tiny{Topological sorting of the components of $\overline{G}_A$}}
    $\ell = | S |$\;
    \tcp{Go through the components in reverse topological order}
    \For{$j = \ell, \ell-1, \ldots, 1$} {
        $H_j = (S_j \cup M, S_j \times M)$  \Comment*[r]{\tiny{Weighted bipartite graph of agents and unallocated items}}
        $w(i,c) = v_i(c)$ for all $i \in S_j$, $c \in M$  \Comment*[r]{\tiny{weights for $H_j$ are the value of the chore for the agent}}
        $N \leftarrow $ maximum weight perfect matching in $H_j$\;
        \For{$i \in S_j$} {
            $A_i \leftarrow A_i \cup \{N(i)\}$\;
            $M \leftarrow M \setminus \{N(i)\}$\;
        }
    }
    \KwRet \text{CwMA}($A, N, M, \V$)
}
\caption{Component-wise Matching Algorithm}
\label{alg:CwMA}
\end{algorithm}

We will show that $\text{CwMA}((\emptyset, \ldots, \emptyset), N, M, \V)$ returns an \EF{1} allocation without generalized envy cycles. Call each recursive call of the algorithm as a \emph{round} of the algorithm. In each round, an agent receives exactly one item.

\begin{restatable}{lemma}{BreakComponent}
\label{lem:BreakComponent}
Let $A$ and $A'$ be the allocations at the start and end of a round respectively. Suppose $S = (S_1, S_2, \ldots S_{\ell})$ was the topological sorting of the components of $\overline{G}_A$. Then $E_{A'}$ has no generalized envy cycles, and for every component $S_{k}'$ in $E_{A'}$, $S_{k}' \subseteq S_j$ for some $j$.
\end{restatable}

\begin{proof}
If $i > j$, then we show that there are no new edges created from an agent in $S_i$ to an agent in $S_j$ during the round. Since $S_i$ comes after $S_j$ in the topological sort, no agent in $S_i$ has an envy or equality edge to any agent in $S_j$ before the round. Thus every agent in $S_i$ strongly prefers their bundle over any bundle in $S_j$. Note that since all agents in $S_i$ pick their items in this round before any agent in $S_j$, they all prefer their new items weakly over any new item allocated to an agent in $S_j$. Thus if $i>j$, there are no new envy or equality edges created from $S_i$ to $S_j$ during the round. This also implies that there cannot be a component $S_k'$ in $E_{A'}$ with vertices from different components of $\overline{G}_A$, because then a new envy or equality edge should have been created from an agent in $S_i$ to an agent in $S_j$ with $i>j$.

We claim that there is no generalized envy cycle created inside $S_j$ during the round. Let $N$ be the maximum weight perfect matching in $H_j$, and let $N(i)$ be the item allocated to agent $i \in S_j$ during this round. Suppose there was a generalized envy cycle $C$ created in $S_j$ after adding the items $\{N(i) \mid i \in S_j \}$. Recall that $i^+$ is the agent in the cycle $C$ that $i$ points to. For all $i \in C$,
\begin{align*}
    v_i(A_i) &\ge v_i(A_{i^+})  \\
    v_i(A_{i^+} \cup \{N(i^+)\}) &\ge v_i(A_i \cup \{N(i)\})
\end{align*}
Putting both of these together, we get that $v_i(N(i^+)) \ge v_i(N(i))$. Since there is an envy edge $(p,p^+)$ in the generalized envy cycle $C$, we get that at least one inequality is strict, $v_p(N(p^+)) > v_p(N(p))$. Then the matching $M$ with $M(i) = N(i)$ if $i \in S_j \setminus C$ and $M(i) = N(i^+)$ if $i \in C$ is a perfect matching with higher weight than $N$, contradicting maximality of $N$.

Thus $E_{A'}$ has no generalized envy cycles, and for every component $S_k'$ in $E_{A'}$, $S_k' \subseteq S_j$ for some $j$.
\end{proof}

\begin{restatable}{lemma}{CwMARoundProperty}
\label{lem:CwMARoundProperty}
Every agent weakly prefers the item she gets in round $t$ over any item any agent gets in round $t+1$.
\end{restatable}

\begin{proof}
This is trivial to see, since if there is an item $c \in M$ that is unallocated that she strongly prefers over the item $c'$ that she obtains in round $t$, then getting matched to $c $ instead of $c'$ gives a matching with higher weight than $N$, a contradiction.
\end{proof}

\begin{restatable}{lemma}{CwMAEF}
\label{lem:CwMAEF}
For an additive instance with indivisible chores, Algorithm~\ref{alg:CwMA} returns an \EF{1} allocation that contains no generalized envy cycles.
\end{restatable}

\begin{proof}
From \Cref{lem:BreakComponent}, we know that the allocation at the end of the last round has no generalized envy cycles. We now show that the allocation is \EF{1}. Since $m$ is a multiple of $n$ (by adding $0$ valued virtual goods if necessary), and since every agent obtains one item in each round, all agents have the same number of items at the end. Denote the number of items each agent has at the end by $\alpha = \frac{m}{n}$. Take any two agents $i$ and $j$. Suppose $i$ envies $j$ in the allocation $A$ obtained using Algorithm~\ref{alg:CwMA}. Then we claim that if we remove the last item $c_i^{\alpha}$ allocated to $i$, $v_i(A_i \setminus \{c_i^{\alpha}\}) \ge v_i(A_j)$. For each round $t$, we know from \Cref{lem:CwMARoundProperty} that $v_i(c_i^t) \ge v_i(c_j^{t+1})$. Thus \[
        v_i(A_i \setminus \{c_i^{\alpha}\}) = \sum_{r=1}^{\alpha-1} v_i(c_i^r)
    \ge \sum_{r=1}^{\alpha-1}v_i(c_j^{r+1}) = v_i(A_j \setminus \{c_j^1\}) \ge v_i(A_j)
\] as required. Thus the allocation is \EF{1}.
\end{proof}

\end{document}